      [1999/12/01 v1.4c Il Nuovo Cimento]
\documentclass{cimento}

\usepackage{graphicx}  
\title{The evolution of the morphological scale of early-type galaxies 
since z=2}
\author{P. Saracco\from{1},
M. Longhetti\from{1},
S. Andreon\from{1},
A. Mignano\from{1}
\thanks{Based on observations made with the NASA/ESA HST obtained at the 
Space Telescope Science Institute.}}
\instlist{\inst{1} INAF - Osservatorio Astronomico di Brera,
Via Brera 28, 20121 Milano, Italy}
\PACSes{\PACSit{98.62}{evolution of galaxies, morphology, scaling relations}
\PACSit{98.52}{elliptical galaxies}
}
\begin{document}

\maketitle

\begin{abstract}

We present the morphological analysis based on HST-NICMOS 
observations in the F160W filter of a sample of 30 early-type galaxies
spectroscopically confirmed at $1.2<z<2$. 
We derive the effective radius R$_e$ and the mean surface brightness 
$\langle\mu\rangle_e$ of galaxies in the rest-frame R-band.
We find that early-types at $z\sim1.5$ are characterized by a surface
brightness much higher then their local counterparts with comparable R$_e$. 
In particular, we find that the mean surface brightness (SB) of these 
early-types should get fainter by $\sim2.5$ mag from $z\sim1.5$ to 
$z\sim0$  to match the SB of the local early-types with comparable R$_e$. 
This evolution exceeds by a factor two the luminosity 
evolution  expected for early-types in this redshift range and more than a 
factor three the one derived from the observed luminosity function of galaxies.
Consequently, an evolution of the effective radius R$_e$ from the epoch of 
their formation towards $z=0$ has to be invoked and the hypothesis of 
fixed size rejected.
\end{abstract}

\section{Introduction}
The formation and the evolution of early-type galaxies (ETGs, elliptical
and bulge-dominated galaxies) occupy an important position among the 
challenges of the observational cosmology.
At least $\sim70\%$ of the stellar mass in the local universe is 
locked into ETGs.
For this reason, the understanding of their build-up and growth is fundamental
to trace the galaxy mass assembly in the Universe.
Evidence for a possible lack of large
early-type galaxies at $z\sim1$ and beyond are coming out
(e.g.\cite{ref:tru06,ref:lo07,ref:tru07}).
At first glance, this could be interpreted as a prove of the merging process
responsible of the growth of local high-mass ellipticals.
On the other hand, the apparently smaller early-types seen at high-redshift 
are characterized by an effective surface brightness brighter than  
the local counterparts, i.e. they are more compact.
However, most of these results are based on
optical observations sampling the blue and UV rest-frame emission of the 
galaxies, particularly sensitive to morphological k-correction and star 
formation episodes, and/or on seeing limited ground-based observations.
Here, we present the morphological analysis based on high-resolution
(FWHM$\sim$0.1 arcsec) HST-NICMOS observations in the F160W filter 
($\lambda\sim1.6$ $\mu$m) of a sample of 30 high-mass ETGs at $1.2<z<2$.

\section{The sample}
The sample of  ETGs we constructed is composed of 
30 galaxies in the redshift range $1.2<z<2$ with
HST-NICMOS observations in the F160W ($\lambda\sim1.6$ $\mu$m) filter.
In particular, images with the NIC2 (0.075 arcsec/pix) camera 
are available for 40\% of the sample and with the NIC3 (0.2 arcsec/pix)  
camera for the remaining 60\%.

The median redshift of the sample is $z_{med}\simeq1.4$.
Ten galaxies come from our own sample of ETGs spectroscopically
confirmed at $1.2<z<1.7$.
The study of their spectro-photometric and morphological properties 
based on multiwavelength data and HST-NICMOS observations
are described in previous works 
\cite{ref:lo05,ref:lo07,ref:sa05}.
The remaining 20 galaxies of the sample have been picked out from 
different surveys on the basis of their morphological classification and 
spectro-photometric properties.
We restricted our selection to those galaxies having both
deep HST-NICMOS observations and spectro-photometric confirmation
of their redshift and spectral type.
On the basis of these criteria we collected 14 ETGs with spectroscopic
confirmation, 10 of which at $1.4<z<1.9$ from  the Galaxy Deep-Deep Survey 
(GDDS, \cite{ref:abr04}) and 4 at $z\sim1.27$ from the sample of 
Stanford et al.\cite{ref:stan97}. 
The remaining 6 ETGs have photometric redshift in the range $1.2<z<2$ 
and they have been selected from the compilation of  
Moriondo et al.\cite{ref:mo00}.


\section{The Kormendy relation and the evolution of the effective radius}
The Kormendy relation (KR, Kormendy 1977) is a linear scaling relation 
between the logarithm of the effective radius R$_e$ [Kpc], i.e. 
the radius containing half of the light, and the mean 
surface brightness $\langle\mu\rangle_e$  [mag/arcsec$^2$]:
{$\langle\mu\rangle_e = \alpha + \beta \log(R_{e})$}

The ETGs follow  this relation with 
a fixed slope $\beta\sim3$ up to $z\sim1$ \cite{ref:dise05}
while the zero point $\alpha$ varies
with the redshift  reflecting the luminosity evolution.
\begin{figure}
\begin{center}
\includegraphics[width=8.5cm]{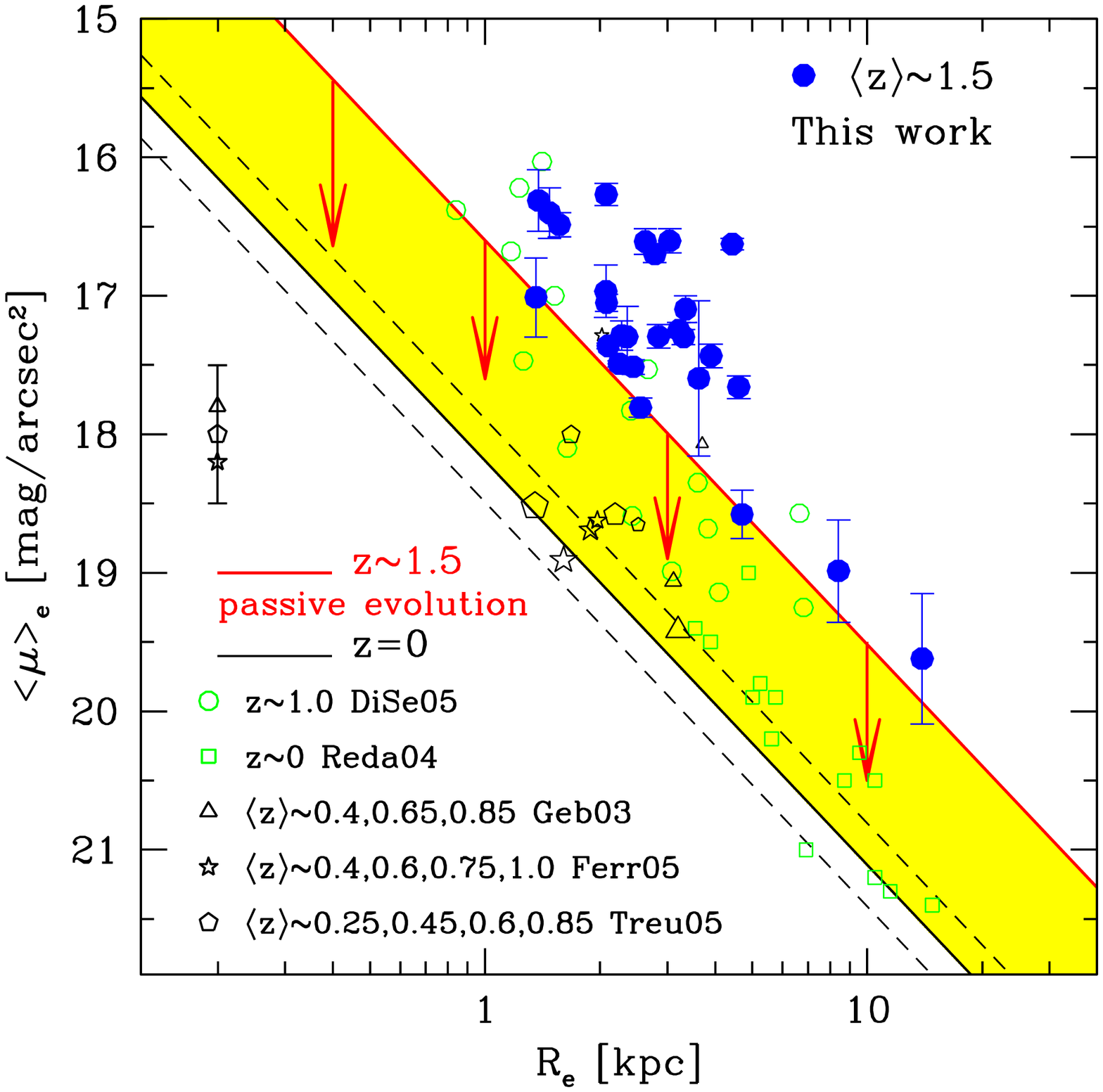}  
\vskip -0.5truecm
\caption{Mean surface brightness 
$\langle\mu\rangle_e$ versus  effective radius R$_e$ [kpc].
The empty symbols are from the literature (see \cite{ref:lo07} for
details).
The thin solid line represents the KR at $z\sim0$ and
the short-dashed lines represent  the $\pm1\sigma$ dispersion of the relation.
The thick (red) line is the KR expected at $z\sim1.5$ in case of PLE.
All the data have been corrected for the cosmological dimming factor 
$(1+z)^4$ thus, any deviation from the KR at $z=0$ reflects 
the evolution of the SB due to the luminosity and/or size evolution 
of galaxies.}
\includegraphics[width=8.5cm]{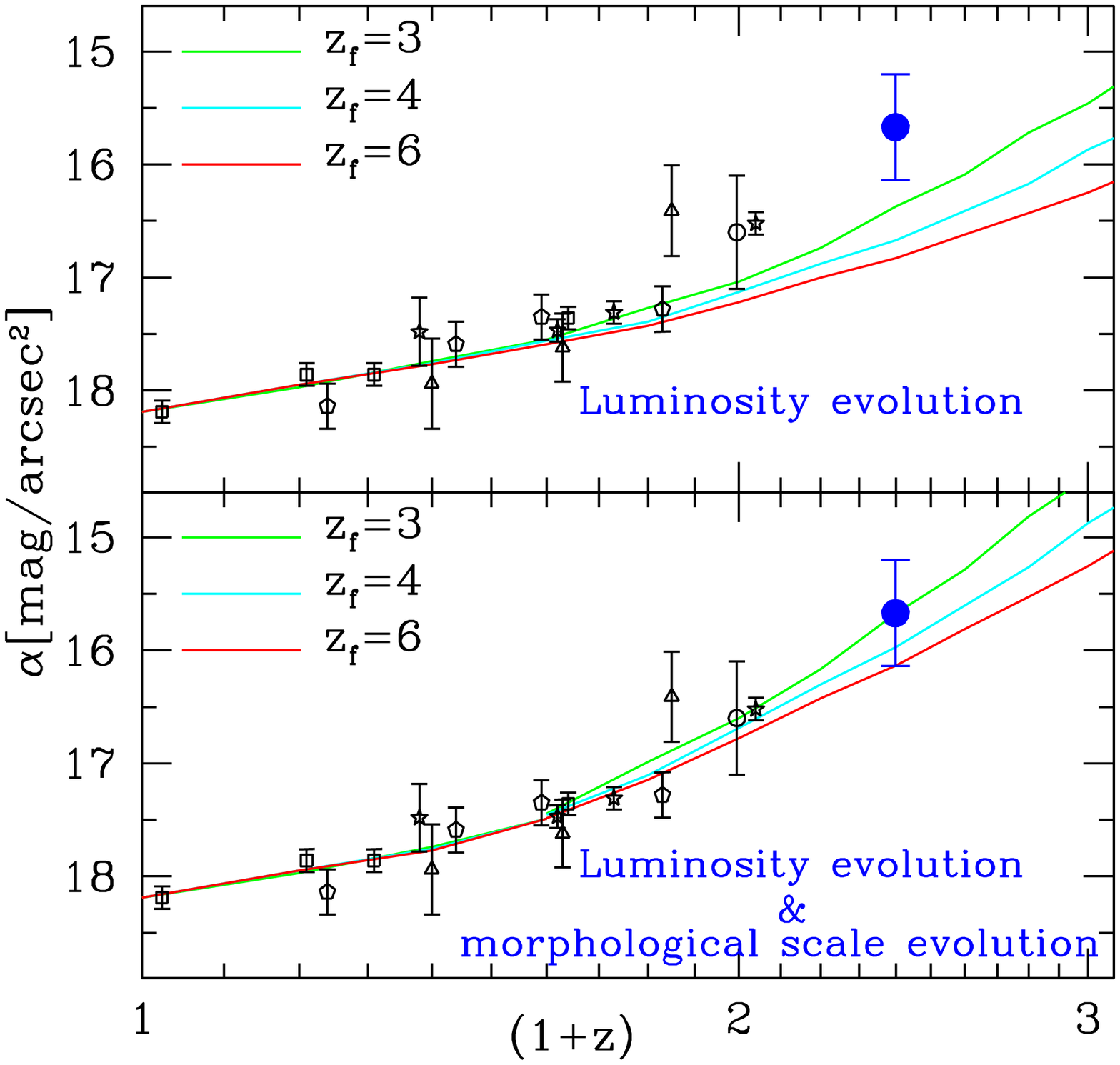} 
\vskip -0.5truecm
\caption{Zero points $\alpha$ of the Kormendy relations 
derived by the various samples as a function of their redshift.
Large filled circle represent the zero point we obtain from our sample.
The lines show the expected evolution of $\alpha$ due to the luminosity 
evolution alone (upper panel) and to the evolution of the morphological scale
R$_e$ in addition to the PLE (lower panel).
Models refer to the same SFH $\tau=0.6$ Gyr starting at the
redshifts of formation (from the top to the bottom) $z_f=3, 4, 6$.}
\end{center}
\end{figure}

We derived the effective radius r$_e$ (arcsec) and the mean surface brightness
(SB) $\langle\mu\rangle_e$ within  r$_e$ of our galaxies by fitting a Sersic 
profile to the observed light profiles. 
The analytic expression of the adopted profiles is
$I(r)=I_{e} exp\{-b_{n}[(r/r_{e})^{1/n}-1]\}$
where $n=4$ and $n=1$ values define the de Vaucouleurs and to the exponential 
(disk) profiles respectively. 
We used galfit \cite{ref:Peng02} to perform the fitting after the 
convolution of the images with the NIC2 and NIC3 PSFs. 
We used simulations \cite{ref:lo07} to asses the reliability of our estimate 
of r$_e$. 
We find that we underestimate  r$_e$ by about 0.05 arcsec in the NIC2 
images and by about 0.03 arcsec in the NIC3 images while the typical rms is
$\sim$0.06 arcsec ($\sim0.5$ Kpc at $z\sim1$).
We derived R$_e$ [kpc] from the values of r$_e$ corrected for the underestimate
 and we
converted the observed SB in the F160W filter into that in the rest-frame R-band using
the proper k-correction for each galaxy (see \cite{ref:lo07} for details).

In Fig. 1 our galaxies (blue filled circles) at $1.2<z<2$ are plotted 
on the [$\mu_{e}$,R$_{e}$] plane
together with those from  lower redshift samples.
The thin (black) solid line is the observed KR in the R band at $z\sim0$
($\langle\mu\rangle_e=18.2+2.92 log(Re)$) while
the thick (red) line is the KR expected at $z=1.5$ 
($\langle\mu\rangle_e=16.6+2.92 log(Re)$) in case of
passive luminosity evolution (PLE). 
These two relations encompass the  (yellow dashed)
region in the  [$\mu_{e}$, R$_{e}$] plane where ETGs at $z\sim1.5$
are expected.
However, all the ETGs of our sample drop out this region.
Their surface brightness exceed by $\Delta\langle\mu\rangle_e\simeq1$ mag 
the one expected in the case of PLE for constant R$_e$. 

In Fig. 2 the values of the zero point $\alpha$ of the KR in the rest-frame 
R band derived from various samples at different redshift are shown 
and compared with the one expected in case of PLE (upper panel).
It is evident the increasing discrepancy between the observed and the 
expected values of $\alpha$ with increasing redshift, confirming
that the luminosity evolution alone cannot reproduce the observed
evolution of the KR from $z\sim1.5$ to $z=0$.
Thus, the other parameter involved in the KR relation, 
the effective radius R$_{e}$,
must evolve and the hypothesis of fixed size must be rejected.
A size evolution of at least a factor 2 from $z\sim1.5$ to $z=0$
 is needed to account for the observed evolution of the KR.
This is shown in the lower panel of Fig. 2 where the expected values of 
$\alpha$ are shown in case of evolution of R$_e$ in addition to the PLE.

\section{Discussion and conclusions}
ETGs at $z\sim1.5$ are characterized by a SB much higher then their 
local counterparts with comparable R$_e$. 
Luminosity evolution is able to account at most only for half of
this excess.
This means that the light of ETGs was actually much more concentrated 
(at least a factor two) in the past then in the present universe, 
i.e. that an evolution of R$_e$ from the epoch of their formation towards 
$z=0$ has occurred. 
However, the appearance of a light profile can change due to both a 
variation of the spatial distribution of the stellar content or  
the presence of color gradients among the stellar populations.
Thus a variation of R$_e$ could coincide both with a variation of the 
stellar density (a contraction of R$_e$ coincides with a contraction of 
the stellar system) or with different star formation histories affecting
the outer and the inner regions of ETGs. 
Both these hypothesis are tightly linked to the formation of ETGs. 
Thus, the understanding of this evolution could open a new window  
in the comprehension of the formation of ETGs and on the assembly of the 
baryonic mass in the universe.

The complete analysis of the size evolution of this sample of ETGs and 
the possible dependence on their stellar mass will be presented in a furthcoming paper
\cite{ref:sar08}.

\acknowledgments
This research has received financial support from the
Istituto Nazionale di Astrofisica (Prin-INAF CRA2006 1.06.08.04)

\end{document}